\documentclass{article}
\usepackage[T1]{fontenc}
\usepackage[utf8]{inputenc}
\usepackage{ismir}
\usepackage{amsmath,cite,url}
\usepackage{graphicx}
\usepackage{color}
\usepackage{amssymb}
\usepackage{bm}
\usepackage{paralist}
\usepackage{booktabs}
\usepackage{pifont}
\usepackage{enumitem}

\usepackage[protrusion=true, expansion=true, tracking=small]{microtype}

\title{CoDiCodec: Unifying Continuous and Discrete Compressed Representations of Audio}

\multauthor
{Marco Pasini$^1$ \hspace{1.5cm} Stefan Lattner$^2$ \hspace{1.5cm} György Fazekas$^1$} {
$^1$Queen Mary University of London, UK \hspace{1cm}
$^2$Sony Computer Science Laboratories, Paris, France\\
{\tt m.pasini@qmul.ac.uk}
}

\def\authorname{M. Pasini, S. Lattner, and G. Fazekas}

\usepackage[bookmarks=false,pdfauthor={\authorname},pdfsubject={\pdfsubject},hidelinks]{hyperref}

\begin{document}

\maketitle


\begin{abstract}
\vspace{-2mm}
Efficiently representing audio signals in a compressed latent space is critical for latent generative modelling. However, existing autoencoders often force a choice between continuous embeddings and discrete tokens. Furthermore, achieving high compression ratios while maintaining audio fidelity remains a challenge.  We introduce CoDiCodec, a novel audio autoencoder that overcomes these limitations by both efficiently encoding global features via \textit{summary embeddings}, and by producing \textit{both} compressed continuous embeddings at \textasciitilde 11 Hz \textit{and} discrete tokens at a rate of 2.38 kbps from the \textit{same} trained model, offering unprecedented flexibility for different downstream generative tasks. This is achieved through Finite Scalar Quantization (FSQ) and a novel \textit{FSQ-dropout} technique, and does not require additional loss terms beyond the single consistency loss used for end-to-end training. CoDiCodec supports both autoregressive decoding and a novel \textit{parallel} decoding strategy, with the latter achieving superior audio quality and faster decoding. CoDiCodec outperforms existing continuous and discrete autoencoders at similar bitrates in terms of reconstruction audio quality. Our work enables a unified approach to audio compression, bridging the gap between continuous and discrete generative modelling paradigms. Pretrained weights are available under [this link].\footnote{\href{https://github.com/SonyCSLParis/codicodec}{https://github.com/SonyCSLParis/codicodec}}
\end{abstract}

\vspace{-2mm}
\section{Introduction}\label{sec:introduction}
\vspace{-1mm}
Efficient, compact audio representations are crucial for applications in Music Information Retrieval (MIR), generative modelling, and compression.  While recent advances in deep learning have demonstrated impressive results in the learning of compressed representations, several key challenges remain.  These include balancing high compression ratios with reconstruction fidelity, enabling both discrete and continuous latent representations for diverse downstream applications, and achieving efficient training and inference without resorting to a complex and unstable training process.

Existing audio autoencoders often fall short in one or more of these areas.  Vector Quantization (VQ)-based approaches, such as SoundStream \cite{zeghidour_soundstream_2022}, EnCodec \cite{defossez_high_2022}, and Descript Audio Codec (DAC, \cite{kumar_high-fidelity_2023}), can excel at high-fidelity reconstruction and are well-suited for training autoregressive language models on the resulting discrete latent tokens \cite{copet_simple_2023, jukebox, agostinelli_musiclm_2023}.  However, their discrete nature makes them less compatible with continuous generative frameworks (e.g., GANs \cite{gan}, diffusion models \cite{diffusion_original, song_score-based_2020, ho_denoising_2020}), as their pre-quantization continuous features are typically high-dimensional and unsuitable for efficient latent modelling.
Continuous autoencoders, such as those used in Mo\^usai \cite{schneider_mousai_2023}, in Musika \cite{pasini_musika_2022, bassnet}, and in the Stable Audio family of generative models \cite{stable_audio, stable_audio_2, stable_audio_open}, address the compatibility issue with continuous latent generative models.  However, they often require multi-stage training procedures, unstable adversarial training objectives, or slow iterative decoding processes.  While Music2Latent \cite{music2latent} introduces a consistency-based autoencoder that achieves single-step decoding and single-loss end-to-end training, it is limited to continuous representations. Furthermore, most continuous autoencoders encode audio into temporally ordered sequences, leading to redundancy by repeatedly encoding global features across embeddings.

This paper introduces CoDiCodec (Continuous-Discrete Codec), a novel audio autoencoder that addresses these limitations. CoDiCodec achieves the following key objectives:

\begin{itemize}[leftmargin=8pt, nosep]
    \item Encoding of both \textbf{compressed continuous embeddings} (\textasciitilde 11 Hz) and \textbf{discrete tokens} (2.38 kbps) of $44.1$ kHz stereo audio from a single model, offering flexibility for downstream tasks without the need for separate models.
    \item Use of \textbf{summary embeddings} \cite{music2latent2} to capture global features, reducing redundancy compared to ordered sequences for better fidelity at similar compression
    \item Leveraging consistency models \cite{song_consistency_2023, improvedconsistencysong}, CoDiCodec is \textbf{trained end-to-end using a single loss}, simplifying the training process and avoiding the complexities of adversarial training or multi-stage procedures.
    \item Support for both autoregressive and a novel, faster \textbf{parallel decoding strategy} for long sequences.
     \item Introduction of \textbf{FSQ-dropout}, enabling higher-quality continuous decoding by bypassing quantization, while promoting informative embeddings suitable for downstream modeling.
    \item An \textbf{improved architecture} designed to increase the proportion of parameters used by the transformer layers compared to convolutional ones, which simplifies the process of scaling, while achieving faster inference speed compared to Music2Latent2.
\end{itemize}

To our knowledge, this is the first work unifying summary embeddings, consistency-based training, and the generation of both continuous and discrete representations from a single audio autoencoder. Our experiments show that CoDiCodec outperforms existing continuous and discrete autoencoders in terms of reconstruction quality measured by FAD \cite{fad} with different backbones. We present comprehensive ablation studies validating the design choices.

\section{Related Work}

\subsection{Audio Autoencoders}

Audio autoencoders aim to learn compressed latent representations of audio signals, typically for dimensionality reduction, generative modeling, or MIR tasks.  These can be broadly divided into those producing discrete and continuous compressed latent representations.

\noindent \textbf{Discrete Latent Representations:} Vector Quantization (VQ \cite{vqvae, vqvae2}) has been a dominant technique for learning discrete audio representations. SoundStream \cite{zeghidour_soundstream_2022}, EnCodec \cite{defossez_high_2022}, and Descript Audio Codec (DAC) \cite{kumar_high-fidelity_2023} use Residual Vector Quantization (RVQ) to achieve high-fidelity audio reconstruction.  These models are particularly well-suited for training autoregressive language models on the resulting discrete tokens \cite{copet_simple_2023, jukebox, agostinelli_musiclm_2023}. However, their discrete nature limits compatibility with continuous generative frameworks, and they often yield lower temporal compression, resulting in longer sequences for downstream tasks compared to continuous methods.


\noindent \textbf{Continuous Latent Representations:}  Several approaches learn continuous latent representations of audio.  The autoencoder used in Musika \cite{pasini_musika_2022} reconstructs both magnitude and phase components of a spectrogram, enabling fast inference. However, it relies on a two-stage training process and an adversarial objective. Mo\^usai \cite{schneider_mousai_2023} uses a diffusion autoencoder, achieving end-to-end training but requiring expensive iterative sampling for decoding. Stable Audio and Stable Audio 2 \cite{stable_audio, stable_audio_2, stable_audio_open} leverage continuous representations to train diffusion-based audio generation models, but the proposed autoencoders still require an objective with multiple adversarial and reconstruction losses. Music2Latent \cite{music2latent} introduces a consistency-based autoencoder, achieving single-step decoding and end-to-end training with a single loss function.  However, it is limited to producing ordered sequences of continuous representations. Music2Latent2 \cite{music2latent2} introduces summary embeddings \cite{titok} that are able to more efficiently encode global features from the input samples, while still encoding to continuous-valued latents.
CoDiCodec, in contrast, can encode both continuous and discrete representations, while still using summary embeddings.

\subsection{Consistency Models}
Consistency models \cite{song_consistency_2023, improvedconsistencysong} represent a class of generative models that enables fast one-step generation. While showing impressive results in image generation \cite{lcm}, their application to audio remains under-explored. CoMoSpeech \cite{comospeech} explores consistency distillation for speech synthesis, requiring a pre-trained teacher. Music2Latent \cite{music2latent} and Music2Latent2 \cite{music2latent2} were the first to use consistency models in an end-to-end audio autoencoder framework.


\begin{figure*}[t]
\centering
\includegraphics[width=\textwidth]{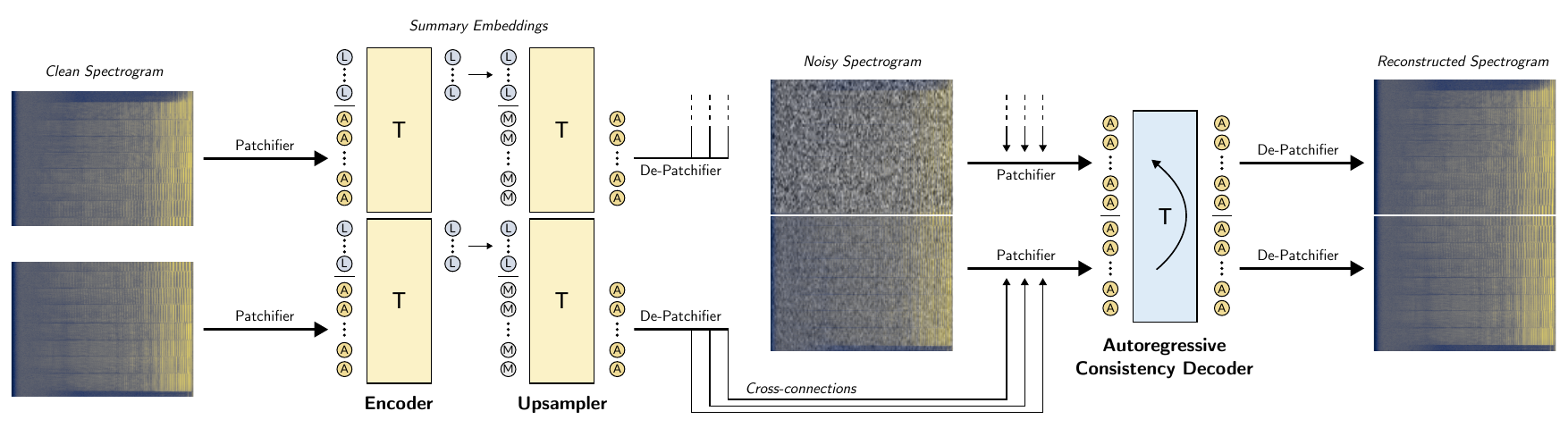}
\caption{Training process. Transformer modules are represented with \textit{T}, audio embeddings with \textit{A}, learned/summary embeddings with \textit{L}, and mask embeddings with \textit{M}. We represent chunked causal masking with a curved arrow.}
\label{fig:architecture}
\end{figure*}

\section{Background}\label{sec:background}

\subsection{Consistency Models}

Consistency models \cite{song_consistency_2023, improvedconsistencysong} are a class of generative models that learn to map any point on a diffusion process trajectory back to the origin of that trajectory. They are based on the probability flow (PF) ordinary differential equation (ODE) \cite{song_denoising_2021}, which describes the evolution of a data sample $x$ perturbed by Gaussian noise with standard deviation $\sigma$:
\begin{equation}
\label{eq:prob_flow_ode}
\frac{dx}{d\sigma} = -\sigma \nabla_x \log p_\sigma(x), \quad \sigma \in [\sigma_{\text{min}}, \sigma_{\text{max}}].
\end{equation}
where $p_\sigma(x)$ is the perturbed data distribution, and $\nabla_x \log p_\sigma(x)$ is the score function. The PF ODE defines trajectories mapping noisy samples $x_\sigma$ to the clean sample $x_{\sigma_{\text{min}}}$ (where $\sigma_{\text{min}} \approx 0$).
Consistency models learn a consistency function $f(x_\sigma, \sigma)$ that directly maps any point on this trajectory to its origin: $f(x_\sigma, \sigma) \mapsto x_{\sigma_{\text{min}}}$, while satisfying the boundary condition $f(x_{\sigma_{\text{min}}}, \sigma_{\text{min}}) = x_{\sigma_{\text{min}}}$.
A consistency model $f_\theta(x_\sigma, \sigma)$ is a neural network parameterized by $\theta$ that approximates the true consistency function. To enforce the boundary condition, consistency models are typically parameterized as $f_\theta(x_\sigma, \sigma) = c_{\text{skip}}(\sigma)x_\sigma + c_{\text{out}}(\sigma)F_\theta(x_\sigma, \sigma)$,
where $F_\theta(x_\sigma, \sigma)$ is a neural network, and $c_{\text{skip}}(\sigma)$ and $c_{\text{out}}(\sigma)$ are chosen such that $c_{\text{skip}}(\sigma_{\text{min}}) = 1$ and $c_{\text{out}}(\sigma_{\text{min}}) = 0$ to satisfy the boundary condition.

\subsection{Consistency Training}
Consistency models can be trained via Consistency Distillation (CD), requiring a pre-trained diffusion model, or Consistency Training (CT). CoDiCodec uses CT, allowing for training in isolation without a pretrained teacher model. In CT, the continuous PF ODE (Eq. \ref{eq:prob_flow_ode}) is discretized using a sequence of noise levels $\sigma_{\text{min}} = \sigma_1 < \sigma_2 < \cdots < \sigma_N = \sigma_{\text{max}}$. The consistency model is trained by minimizing the following loss:
\begin{equation}
\label{eq:consistency_training_loss}
\mathcal{L}_{\text{CT}} = \mathbb{E} \left[ \lambda(\sigma_i, \sigma_{i+1}) d\left(f_\theta(x_{\sigma_{i+1}}, \sigma_{i+1}), f_{\theta^-}(x_{\sigma_i}, \sigma_i)\right) \right],
\end{equation}
where $x \sim p_{\text{data}}$ is a training sample, $\sigma_i$ and $\sigma_{i+1}$ are adjacent noise levels, $x_{\sigma_i}$ and $x_{\sigma_{i+1}}$ are corresponding noisy versions of $x$, $d(x, y)$ is a distance metric, and $\lambda(\sigma_i, \sigma_{i+1})$ is a weighting function.  $f_\theta$ is the student model, and $f_{\theta^-}$ is the teacher, with parameters $\theta^- \gets \text{stopgrad}(\theta)$. The loss minimizes the distance between model outputs at adjacent noise steps $\sigma_i, \sigma_{i+1}$, using the teacher $f_{\theta^-}$ to provide the targets.
Post-training, generation from noise $x_{\sigma_{\text{max}}}$ can occur in one step ($x = f_\theta(x_{\sigma_{\text{max}}}, \sigma_{\text{max}})$) where $x_{\sigma_{\text{max}}} \sim \mathcal{N}(0, \sigma_{\text{max}}^2 I)$, or multiple steps.

\subsection{Finite Scalar Quantization (FSQ)}
Finite Scalar Quantization (FSQ) \cite{fsq} is a simple quantization technique and, unlike Vector Quantization (VQ \cite{vqvae,vqvae2}), it does not require additional loss terms. It is also shown to achieve almost perfect codebook utilization even with large codebook sizes. FSQ bounds a value $x$ in $[-N, N]$ where $N$ is integer, rounds it, and rescales:
\begin{equation}
\hat{x} = \frac{\text{round}(N \cdot \tanh(x))}{N},
\end{equation}
where $\hat{x}$ is the quantized value, and $\text{round}(\cdot)$ denotes the rounding operation. Applied element-wise to a $D$-dimensional vector $\mathbf{x}$, each element $\hat{x}_i$ takes one of $2N+1$ discrete values in $[-1, 1]$, yielding an implicit codebook of size $(2N+1)^D$.
The gradient of the non-differentiable rounding operation is approximated using the straight-through estimator \cite{ste}.

\section{CoDiCodec}
Following previous work \cite{drumgan, comparing, music2latent, music2latent2}, CoDiCodec operates on complex Short-Time Fourier Transform (STFT) spectrograms. 
To address the skewed distribution of different frequency bins, we apply an amplitude transformation \cite{speech_enhancement}: $\tilde{c} = \beta |c|^\alpha e^{i \angle(c)}$, where $c$ and $\tilde{c}$ are the original and transformed STFT coefficients, $\alpha \in (0, 1]$ is a compression exponent that emphasizes lower-energy components, $\angle(c)$ is the phase angle of $c$, and $\beta \in \mathbb{R}^{+}$ is a scaling factor. We treat the complex spectrogram as a two-channel (real/imaginary) representation.
\vspace{-2mm}
\subsection{Architecture}
\vspace{-2mm}
The proposed architecture (Fig. \ref{fig:architecture}) consists of an encoder, an upsampler, and a consistency model decoder. The model operates on pairs of consecutive audio chunks.
\noindent \textbf{Encoder}: It takes a spectrogram chunk $x \in \mathbb{R}^{C \times F \times T}$ ($C=2 \times \text{channels}$, $F$ and $T$ are the number of frequency bins and time frames) and downsamples it via a convolutional patchifier. The flattened features (audio embeddings) are concatenated with $K$ learnable \textit{summary embeddings} and fed into transformer blocks \cite{transformer} (T in Fig. \ref{fig:architecture}) for summary embeddings to gather global context. Only the $K$ summary embeddings are retained, projected to $d_{lat}$, and processed via $\tanh$ (for continuous output) or FSQ (for discrete tokens converted to indices at inference).

\noindent \textbf{Upsampler}: It mirrors the encoder structure but upsamples instead of downsampling. It takes $K$ summary embeddings (discrete tokens are mapped back to vectors), concatenates learnable mask embeddings, and processes them through transformer blocks to ``de-compress'' information from the summary embeddings. The resulting audio embeddings are reshaped and upsampled by a convolutional de-patchifier. Its sole purpose is providing intermediate feature maps as \textit{cross-connections} to the decoder: since the consistency model decoder generates samples in one step, it is crucial to provide information about which sample to reconstruct to the first layers of the decoder \cite{music2latent}.

\noindent \textbf{Consistency Decoder}: It is trained to map a noisy spectrogram $x_\sigma$ to a clean one, conditioned on upsampler cross-connections. A patchifier downsamples the input noisy spectrogram $x_\sigma$. Cross-connections from the upsampler are added to feature maps at each resolution level: this is possible because of the exact symmetry of the patchifier with respect to the de-patchifier of the upsampler. The output is flattened and fed into a stack of transformer blocks. Crucially, transformers operate on consecutive chunk pairs ($x_{\sigma, \text{left}}, x_{\sigma, \text{right}}$) with chunked causal masking (right chunk attends to left, not vice-versa) to enable autoregressive decoding. A de-patchifier upsamples the output to the original spectrogram dimension. Skip connections additively combine the feature maps from the patchifier to the corresponding ones in the de-patchifier. The forward pass is:
\begin{align*}
    \hat{x}_{\text{left}},\hat{x}_{\text{right}} &= 
    \text{Dec}_{\sigma_{\text{left}},\sigma_{\text{right}}}(\text{Up}(\text{Enc}(x_{\text{left}})), x_{\text{left}}+\sigma_{\text{left}}\varepsilon_{\text{left}}, \\
    &\qquad \qquad \qquad \text{Up}(\text{Enc}(x_{\text{right}})), x_{\text{right}}+\sigma_{\text{right}}\varepsilon_{\text{right}})
\end{align*}
where Enc, Up, and Dec are the Encoder, Upsampler, and Decoder. $\varepsilon \sim \mathcal{N}(0, I)$ and noise levels $\sigma$ are sampled independently. End-to-end training uses the consistency loss \cite{improvedconsistencysong}:
\begin{equation*}
    \mathcal{L} = \mathbb{E} \left[ \frac{1}{\Delta\sigma} d\left(\text{Dec}_{\sigma_{\text{left}}+\Delta\sigma,\sigma_{\text{right}}+\Delta\sigma}, \text{sg}\left(\text{Dec}_{\sigma_{\text{left}},\sigma_{\text{right}}}\right)\right) \right]
\end{equation*}
with Pseudo-Huber distance $d(\cdot)$, step $\Delta\sigma$, and stop-gradient \textit{sg}. We use the EDM parameterization \cite{karras_elucidating_2022, song_consistency_2023}, continuous log-normal noise sampling \cite{karras_elucidating_2022}, and an exponential $\Delta\sigma$ schedule \cite{music2latent, easy_consistency_models}.

\begin{figure}[t]
\centering
\begin{tabular}{cc}
\includegraphics[width=0.45\columnwidth]{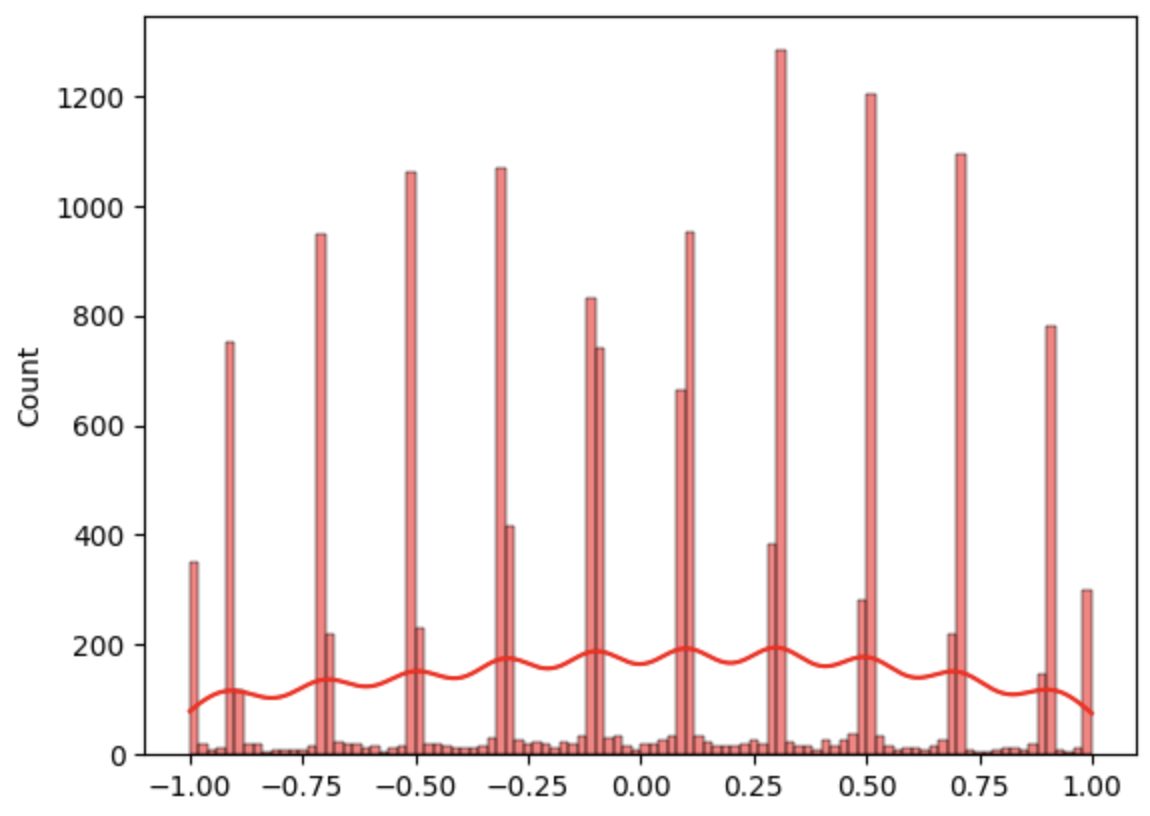} &
\includegraphics[width=0.45\columnwidth]{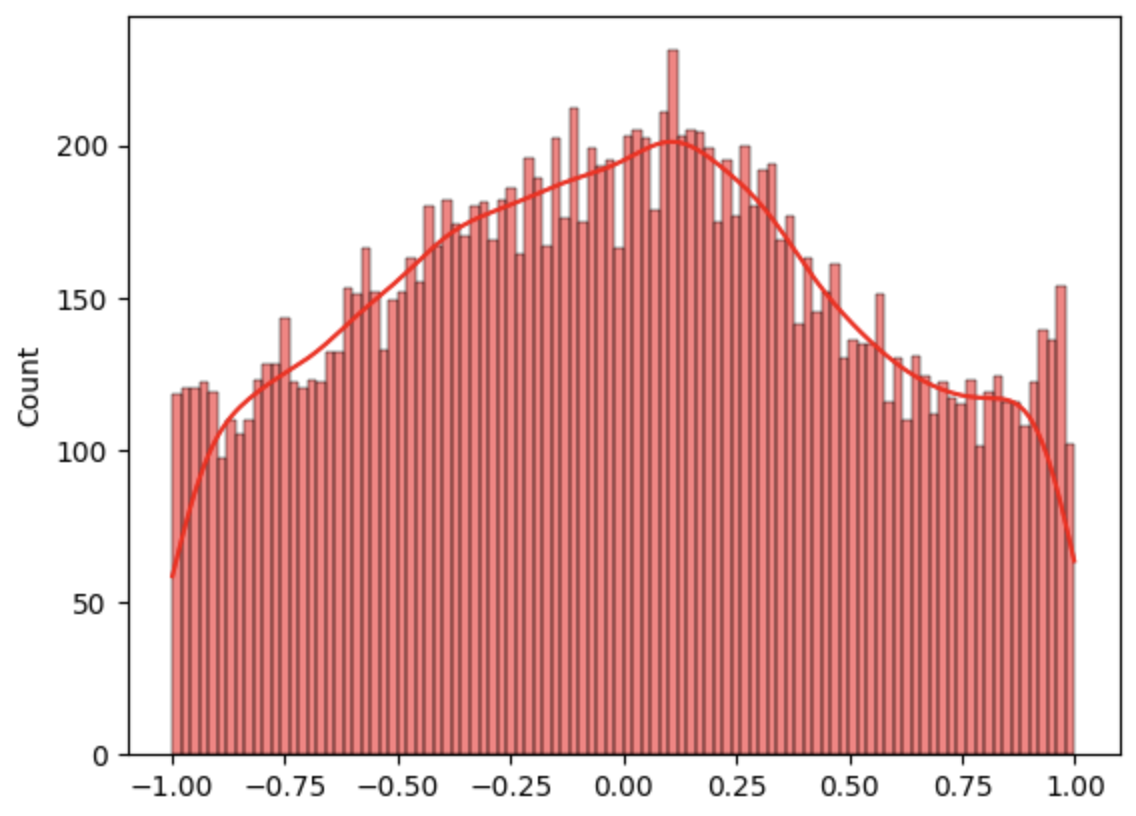} \\
(a) Standard FSQ & (b) FSQ-dropout p=0.75
\end{tabular}
\caption{Distribution of continuous latent embeddings of an evaluation audio sample before the rounding operation (a) with standard FSQ, and (b) with FSQ-dropout with p=0.75. FSQ-dropout encourages a more uniform distribution, utilizing the full range between -1 and 1.}
\label{fig:fsq_distribution}
\end{figure}

\noindent \textbf{FSQ-dropout}: To enable decoding from both discrete FSQ tokens and more expressive continuous embeddings using the same model, we introduce \textit{FSQ-dropout}. Standard FSQ training causes continuous pre-quantization values ($\tanh(\mathbf{z})$) to cluster near quantization levels (Fig. \ref{fig:fsq_distribution}(a)), limiting expressiveness. Even if the encoder did produce a more uniform distribution of continuous values, we would be forced to apply the FSQ rounding operation before decoding, thus rounding away the additional information, since during training FSQ is always enabled. FSQ-dropout addresses this: during training, with probability $p$, we bypass FSQ's rounding step, feeding the continuous $\tanh(\mathbf{z})$ directly to the upsampler; otherwise, we apply standard FSQ rounding:
\begin{equation}
\tilde{\mathbf{z}} =
\begin{cases}
\tanh(\mathbf{z}), & \text{with probability } p \\
\frac{\text{round}(N \cdot \tanh(\mathbf{z}))}{N}, & \text{with probability } 1-p
\end{cases}
\end{equation}
where choosing $N$ results in $2N + 1$ FSQ quantization levels. This encourages the encoder to produce more informative continuous embeddings across the full $[-1, 1]$ range (Fig. \ref{fig:fsq_distribution}(b)) and trains the decoder to accept both discrete and continuous inputs, enabling higher-fidelity continuous reconstruction at inference. We note that a similar technique is proposed in \cite{stable_codec}, using a combination of FSQ and uniform noise dithering.

\noindent \textbf{Random Mixing}: We also introduce \textit{random mixing} as a data augmentation technique. With a probability of 0.5, two randomly selected training samples are mixed (added together) to create a new training sample. This encourages the model to be robust to complex audio scenes with multiple sources. We ablate the effectiveness of this technique in Section \ref{sec:experiments}.
\vspace{-2mm}
\subsection{Decoding Process}
\label{sec:decoding}
\vspace{-2mm}
CoDiCodec supports two decoding strategies: autoregressive decoding, and a novel \textit{parallel decoding} strategy.

\noindent \textbf{Autoregressive Decoding}: Autoregressive decoding is well-suited for interactive applications requiring low latency.  In this mode, CoDiCodec generates audio sequentially, chunk by chunk, conditioning the generation of each new chunk on the previously decoded one. For a detailed formalization, we refer the reader to the Music2Latent2 paper \cite{music2latent2}.


\noindent \textbf{Parallel Decoding}: While autoregressive decoding is suitable for interactive applications, it can be inefficient for decoding long sequences, as each chunk must be processed sequentially. We introduce a novel \textit{parallel decoding} strategy that addresses this limitation. 

At a high level, we decode adjacent pairs of compressed latents in parallel, and shift the pairs by one at each denoising step to avoid boundary artifacts.
More specifically, given a sequence of $T$ sets of summary embeddings, each set encoding information about an audio chunk, we split them into $\lceil T/2 \rceil$ pairs.  If $T$ is odd, the last set is paired with a set of zeroed-out summary embeddings.  Each pair of summary embeddings is then processed independently. 
The decoding process involves multiple denoising steps ($S$).
    \textit{Step 1}:  Each pair of summary embeddings is decoded by the consistency model in parallel, starting from pure noise representations for both the left and right chunks.
    \textit{Step} $s$ ($1 < s \leq S$):  The previously decoded chunks are concatenated, and the \textit{pairs are shifted by one position}.  For example, if chunks 0 and 1 were paired in the previous step, chunks 1 and 2 are paired in the current step.  Gaussian noise with a decreasing standard deviation $\sigma_{\text{cond}, s}$ is added to all chunks. The consistency model then denoises each pair of chunks, conditioned on the corresponding summary embeddings. A linearly decreasing noise schedule ensures that the model gradually refines the decoded audio samples.

This iterative process, with shifting pairs, effectively allows information to propagate across the sequence, mitigating boundary artifacts that would arise from independently decoding fixed pairs. The number of steps, $S$, controls the trade-off between computational cost and reconstruction quality. While the memory usage of autoregressive decoding is constant regardless of the length of the sequence, for parallel decoding it scales linearly with length (number of chunks), since the model performs multiple decoding steps at the same time.

\vspace{-2mm}
\subsection{Implementation Details}\label{subsec:implementation}
\vspace{-2mm}
\noindent \textbf{Architecture}: CoDiCodec features a scaled-up architecture compared to Music2Latent2 \cite{music2latent2}, prioritizing transformer blocks over convolutional layers for ease of scalability \cite{scaling, chinchilla_scaling}. The STFT representation uses hop=1024, compared to 512 in Music2Latent, and window=2048. The convolutional patchifiers and de-patchifiers have 5 resolution levels, compared to 7 in Music2Latent2.  We use [3, 3, 3, 1] convolutional layers per level, and [64, 128, 256, 512] channels per level. Downsampling/upsampling are performed 3 times, with a factor of 2 along both time and frequency, except for the middle level, where only the frequency axis is downsampled/upsampled by a factor of 4.  The encoder, upsampler, and consistency model each have 12 transformer blocks.  These blocks have a hidden\_dim=512 (compared to 256 in Music2Latent2), head\_dim=128, and mlp\_mult=4.  For each input chunk, the encoder produces $K=128$ summary embeddings, each with a dimensionality of $d_{lat} = 4$. We can then reshape them to 8 embeddings with 64 channels (resulting in \textasciitilde 11 Hz representations for stereo 44.1 kHz audio). Since they are not a temporally ordered sequence, they can be freely reshaped for different time-dimension vs. channels trade-offs. Noise levels ($\sigma$) are encoded using sinusoidal embeddings \cite{transformer} with 512 channels.
Training uses audio samples of 67,072 samples (approximately 1.5 seconds at 44.1 kHz), with STFT spectrograms split into two consecutive 32-frame chunks. We use a batch size of 20 and train for 2 million iterations. We use RAdam \cite{radam} with a learning rate of $1 \times 10^{-4}$, $\beta_1 = 0.9$, and $\beta_2 = 0.999$. A cosine learning rate decay is applied, reaching a final learning rate of $0$. An Exponential Moving Average (EMA) of the model parameters is maintained with a momentum of 0.9999. For FSQ, we use $N=5$, resulting in 11 quantization levels per dimension and an implicit codebook size of $11^{4}=14'641$, which is much lower than what modern LLMs \cite{llama, llama2, llama3} use. Given the 128 tokens per chunk, this results in a 2.38 kbps rate for stereo 44.1 kHz audio.  FSQ-dropout is used with $p=0.75$, following our ablation results. We use the consistency training framework of \cite{music2latent}, with an initial consistency step of $\Delta t_0 = 0.1$ and a final exponent of $e_K = 2$. Random mixing data augmentation is applied with a probability of 0.5.  Training is performed on a single A100 GPU and takes \textasciitilde two weeks. The model has \textasciitilde 150 million parameters.

\vspace{-3mm}
\section{Experiments and Results}
\label{sec:experiments}
\vspace{-2mm}
\noindent \textbf{Data}: We train CoDiCodec on a combination of three datasets:
MTG-Jamendo \cite{mtg_jamendo} for music (~3k hours), the speech (~800 hours) and general audio (~200 hours) samples from DNS Challenge 4 \cite{dns}, and M4singer \cite{m4singer} for singing voice (~30 hours). We sample the training datasets with weights $[4, 1.5, 4, 1]$, respectively, during training. We choose these weights in order to train CoDiCodec to be robust to speech and general audio, while still focusing on music. Since we are mainly interested in the performance of our model on musical samples, we use MusicCaps \cite{agostinelli_musiclm_2023} as the evaluation dataset. We manually verify that none of the samples in MusicCaps are present in the training sets. 

\noindent \textbf{Baselines}: For continuous representation baselines, we include: Musika \cite{pasini_musika_2022}, an autoencoder reconstructing magnitude and phase spectrograms; LatMusic \cite{bassnet}, an autoencoder designed for latent diffusion models in music accompaniment generation; Mo\^usai \cite{schneider_mousai_2023}, which provides two diffusion autoencoder models (v2 and v3) with differing compression ratios; Music2Latent \cite{music2latent} and Music2Latent2 \cite{music2latent2}, two consistency-based autoencoders; and the autoencoder used in Stable Audio Open \cite{stable_audio_2, stable_audio_open}. All these models have compression ratios from 32x to 128x, calculated as waveform values in divided by latent values out. We also include Descript Audio Codec (DAC) \cite{kumar_high-fidelity_2023}, a high-fidelity RVQ-based autoencoder producing discrete representations, using both its 2.67 kbit/s and 8 kbit/s configurations.

\noindent \textbf{Metrics}: We use:
\textit{SI-SDR} (Scale-Invariant Signal-to-Distortion Ratio) \cite{sisdr}, which measures the distance between the reconstructed and original waveforms; \textit{ViSQOL} (Virtual Speech Quality Objective Listener) \cite{visqol, visqolaudio, visqolv3}, which estimates pair-wise perceptual audio quality, providing a MOS-like score; \textit{FAD} (Fréchet Audio Distance) \cite{fad}, which measures the distance between the distributions of real and generated audio features from a pretrained VGGish \cite{vggish}, assessing overall audio quality; \textit{FAD}\_\textit{clap}, a variant of FAD that uses CLAP \cite{clap} features, shown to better correlate with human perception of audio quality \cite{fad_correlation}.
\begin{table}[h]
\begin{small}
\centering
\begin{tabular}{lcccccc}
\toprule
& \multicolumn{2}{c}{Continuous} & \multicolumn{2}{c}{Discrete} \\
\cmidrule(lr){2-3} \cmidrule(lr){4-5}
Model & $\text{FAD}_\text{clap}\downarrow$ & $\text{FAD}\downarrow$ & $\text{FAD}_\text{clap}\downarrow$ & $\text{FAD}\downarrow$ \\
\midrule
M2L2          & 0.0218 & 0.784 & - & - \\
+ mix aug.          & 0.0208 & 0.745 & - & - \\
+ new arch.        & 0.0178      & 0.635      & -          & -    \\
+ 128 lat.      & 0.0154      & 0.568     & -          & -       \\
+ FSQ        & -     & -        & 0.0182            & 0.704        \\
d.o. p=0.25        & 0.0173     & 0.628         & 0.0184            & 0.725        \\
d.o. p=0.5        & 0.0169      & 0.618        & 0.0191            & 0.718      \\
d.o. p=0.75        & 0.0161      & 0.599        & 0.0187            & 0.716    \\
\bottomrule
\end{tabular}
\caption{Incremental ablation study.}
\label{tab:incremental_ablation}
\end{small}
\end{table}
\vspace{-2mm}
\subsection{Ablation Study}\label{sec:incremental_ablation}
\vspace{-2mm}
We conduct an ablation study to validate the key design choices of CoDiCodec. We train all ablated models for 400k iterations with a batch size of 20, keeping other training parameters and dataset consistent with the full model. We start by evaluating the same architecture presented in Music2Latent2. We then incrementally add changes to individually evaluate their effect. We first use the random mixing augmentation, then change to our proposed architecture, then use 128 4-dimensional summary embeddings instead of 8 64-dimensional summary embeddings (both having the same total dimensionality and resulting compression ratio), and finally use FSQ-dropout with varying values of the dropout probability $p$.
For each configuration, we report FAD and $\text{FAD}_\text{clap}$ for both continuous embeddings and discrete tokens, when applicable. Table \ref{tab:incremental_ablation} shows how using the random mixing augmentation, changing to our proposed architecture, and re-distributing the same latent space dimensionality from 8 latents with 64 channels to 128 latents with 4 channels, all independently contribute to lower $\text{FAD}_{\text{clap}}$ and FAD. Introducing FSQ performs slightly worse (as expected, due to quantization), but enables discrete tokens. FSQ-dropout with $p=0.75$ allows us to both recover a similar discrete tokens performance as the standard FSQ variant, and similar continuous embeddings performance as the fully continuous variant. We thus use this configuration for the remaining experiments.
\begin{table*}[h]
\begin{small}
\centering
\begin{tabular}{lccccccrr}
\toprule
Model        & Stereo & Representation & Compression Ratio & Bitrate & SI-SDR $\uparrow$ & ViSQOL $\uparrow$ & $\text{FAD}_\text{clap}\downarrow$ & $\text{FAD}\downarrow$ \\
\midrule
Musika       &  \ding{55}  & Continuous & 64x & -     & -25.81 & 3.80 & 0.103 & 2.308      \\
LatMusic     &  \ding{55}   & Continuous  & 64x   & -    & -27.32 & 3.95 & 0.050 & 1.630      \\
Mo\^usai\_v2 &  \ding{51}   & Continuous  & 64x   & -    & -21.44 & 2.36 & 0.731 & 4.687    \\
Mo\^usai\_v3  & \ding{51}   & Continuous  & 32x    & -   & -17.47 & 2.28 & 0.647 & 4.473       \\
Music2Latent &  \ding{55}   & Continuous  & 64x     & -  & -3.85 & 3.84 & 0.036 & 1.176       \\
Music2Latent2 &  \ding{51}    & Continuous  & \textbf{128x}  & -     & -2.29 & 3.91 & 0.023 & 0.717       \\
Stable Audio & \ding{51} & Continuous  & 64x & - & \textbf{6.04} & \textbf{4.08} & 0.107 & 1.017      \\
\midrule
CoDiCodec (AR) & \ding{51} & Continuous & \textbf{128x} & -   & -0.28 & 3.95 & 0.0120  & 0.390 \\
CoDiCodec (Par., s=3)    &  \ding{51}  & Continuous & \textbf{128x} & - & -0.08 & 3.94 & 0.0114      &0.355   \\
CoDiCodec (Par., s=4)    &  \ding{51}   & Continuous & \textbf{128x} & - & -0.01 & 3.95 & \textbf{0.0112}      &\textbf{0.344}   \\
\midrule
DAC  &  \ding{55}        & Discrete & - & 2.67 kbps & 2.80      & 3.87        & 0.174            & 3.791        \\
DAC &  \ding{55}        & Discrete  & - & 8 kbps & \textbf{9.48} & \textbf{4.21} & 0.041 & 0.966        \\
\midrule
CoDiCodec (AR) & \ding{51} & Discrete & - & \textbf{2.38 kbps}  & -0.95 & 3.89 & 0.0136  & 0.485 \\
CoDiCodec (Par., s=3)    &  \ding{51}  & Discrete & -& \textbf{2.38 kbps}   & -0.74 & 3.88 & 0.0130      &0.431   \\
CoDiCodec (Par., s=4)    &  \ding{51}   & Discrete & - & \textbf{2.38 kbps} & -0.66 & 3.90 & \textbf{0.0127}      &\textbf{0.427}   \\
\bottomrule
\end{tabular}
\caption{Audio quality and reconstruction metrics.}
\label{tab:main_results_audio}
\end{small}
\end{table*}
\begin{figure}[h]
\centering
\includegraphics[width=\columnwidth]{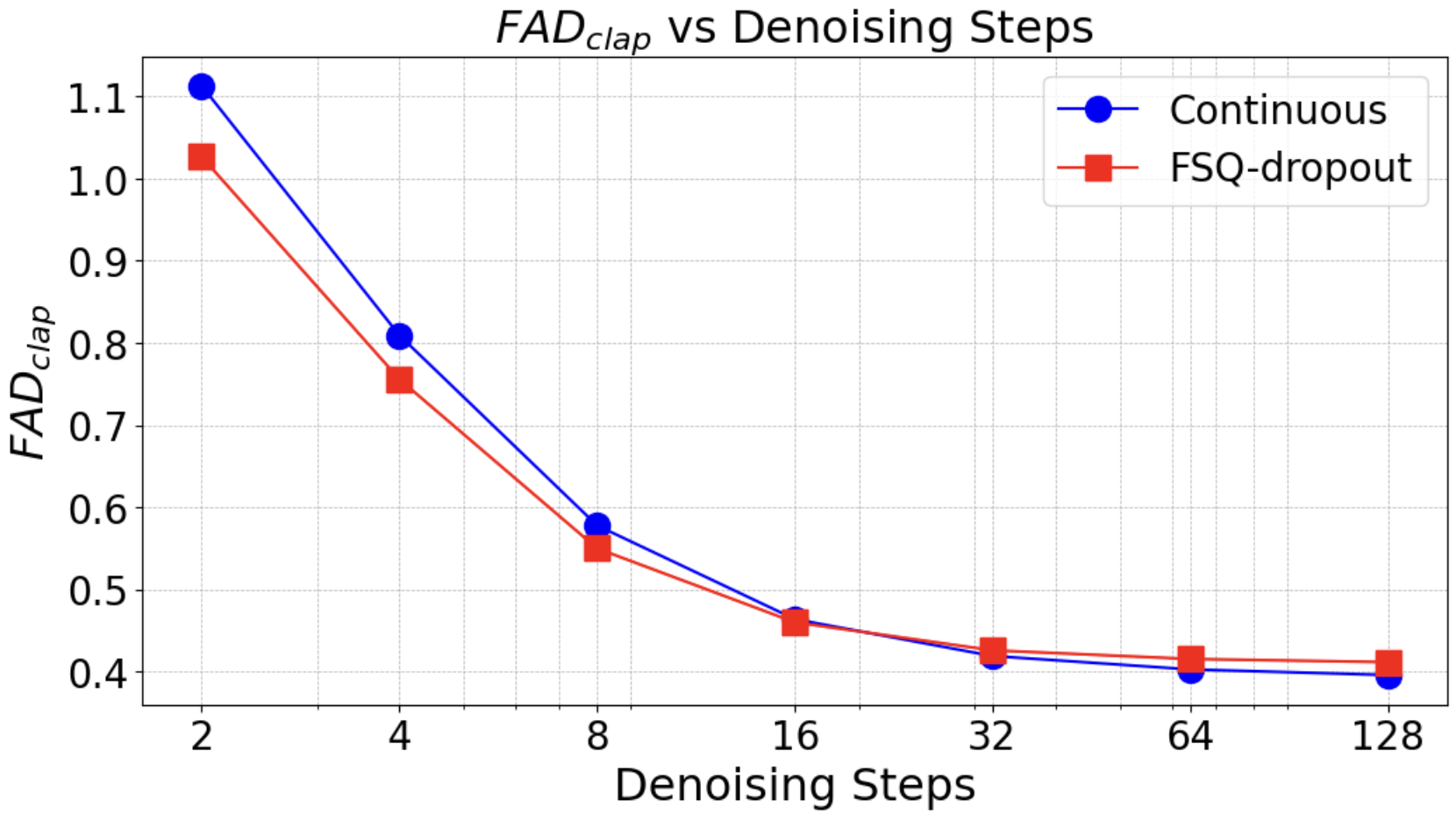}
\caption{Downstream generative modeling $\text{FAD}_\text{clap}$ with respect to number of denoising steps.}
\label{fig:downstream_gen}
\end{figure}
\vspace{-5mm}
\subsection{Downstream Generative Modeling}
\vspace{-2mm}
To assess the impact of the introduced compressed embeddings FSQ-based constraint on downstream generative modeling, we train unconditional generative models using Rectified Flow \cite{rectified_flow} on continuous latent representations from two configurations:

1.  \textit{Continuous}:  Embeddings from the ``+ 128 lat.'' model from the ablation study (Section \ref{sec:incremental_ablation}), which does not use any FSQ, but a simple tanh bottleneck. The distribution of the latent values follows a gaussian-like distribution, which we scale to have unit standard deviation for the training data.

2.  \textit{FSQ-dropout}: Embeddings from the ``d.o. p=0.75'' model, taken without the FSQ rounding operation. In this case, we first apply an $\text{atanh}$ operation to project the FSQ-dropout continuous values from a uniform (Fig. \ref{fig:fsq_distribution}(b)) into a comparable gaussian-resembling distribution, and then rescale them to have unit standard deviation.

For each setting, we train a \textasciitilde 100M parameter Rectified Flow DiT \cite{dit} for 200k iterations with a batch size of 128, using latents of 10-second samples.  We use an internal dataset of ~100k single instrument sources as training data. We then generate 1000 samples and evaluate them using $\text{FAD}_\text{clap}$, varying the number of DiT denoising steps during generation. In Fig. \ref{fig:downstream_gen} we show that while both configurations converge to a comparable $\text{FAD}_\text{clap}$ with a large number of denoising steps, the model trained on FSQ-dropout embeddings (Setting 2) achieves slightly lower $\text{FAD}_\text{clap}$ when using less than 32 denoising steps. We hypothesize that the implicit regularization provided by FSQ-dropout can be beneficial for latent generative modelling: the decoder appears to be slightly more ``robust'' to noisy generations of the downstream model. We will further investigate this hypothesis in future work.
\begin{table}[h]
\begin{small}
\centering
\begin{tabular}{lcc}
\toprule
Model        & Encoding (s) & Decoding (s) \\
\midrule
Music2Latent2 (AR)      & 0.44 & 4.53  \\
Ours (AR)      & 0.34 & 3.22  \\
Ours (Par. s=3) & 0.34 & 2.23  \\
Ours (Par. s=4) & 0.34 & 2.89  \\
Ours (Par. s=5) & 0.34 & 3.51  \\
\bottomrule
\end{tabular}
\caption{Inference speed comparison (60-second audio).}
\label{tab:inference_speed}
\end{small}
\end{table}
\vspace{-5mm}
\subsection{Audio Quality and Reconstruction}
\vspace{-1mm}
We evaluate CoDiCodec trained as described in Sec. \ref{subsec:implementation}.
Table \ref{tab:main_results_audio} presents the audio quality and reconstruction accuracy results.  We evaluate both autoregressive (AR) and parallel (Par. using 3 and 4 denoising steps) decoding. We also evaluate both continuous (Cont.) and discrete (Disc.) representations. CoDiCodec significantly outperforms all continuous autoencoder baselines in terms of FAD and FAD\_{clap}. While some baslines achieve higher SI-SDR and ViSQOL, they are explicitly trained with reconstruction losses, while CoDiCodec only uses a generative loss: general audio quality is thus prioritised over reconstruction of the exact same signal, which hurts these pairwise metrics. Crucially, the proposed parallel decoding strategy achieves the best audio quality results, for both continuous and discrete representations. We provide samples here\footnote{\href{https://sonycslparis.github.io/codicodec-companion/}{sonycslparis.github.io/codicodec}}.
\vspace{-2mm}
\subsection{Inference Speed}
\vspace{-1mm}
We measure the encoding and decoding speed by processing a 60-second audio sample on a single RTX 3090 GPU.
Table \ref{tab:inference_speed} shows that CoDiCodec achieves faster encoding than Music2Latent2, and also substantially faster decoding using the exact same autoregressive decoding strategy. Parallel decoding can further provide lower times if using less than 5 steps. Assuming unlimited memory at our disposal for parallel processing, decoding even longer samples would inevitably widen the gap.
\vspace{-3mm}
\section{Conclusion}
\vspace{-2mm}
This paper introduced a novel audio autoencoder producing both continuous embeddings and discrete tokens from a single model, trained end-to-end with a single consistency loss. This is achieved via finite scalar quantization and our proposed FSQ-dropout technique, which allows for expressive continuous latents that perform well for downstream generative modelling. CoDiCodec leverages summary embeddings for high compression and supports both autoregressive and a novel, faster parallel decoding strategy, outperforming existing autoencoders in audio quality metrics. A novel architecture is designed for scalability, focusing on transformer layers. Future work will explore scaling up the model, applying it to diverse audio domains, and investigating its representations for a broader range of MIR tasks, fully exploring the potential of unifying compressed continuous and discrete representations under a single model.

\section{Acknowledgements}
This work is supported by the EPSRC UKRI Centre for
Doctoral Training in Artificial Intelligence and Music (EP/S022694/1) and Sony Computer Science Laboratories Paris.

\bibliography{ISMIRtemplate}

%
%
%
%

\end{document}